# Quantum-noise-limited optical frequency comb spectroscopy

Aleksandra Foltynowicz[*], Ticijana Ban, Piotr Masłowski, Florian Adler and Jun Ye

*JILA, National Institute of Standards and Technology and University of Colorado,*

*Department of Physics, University of Colorado, Boulder, Colorado, USA*

[*] aleksandra.matyba@jila.colorado.edu

We achieve a quantum-noise-limited absorption sensitivity of $1.7 \times 10^{-12}$ cm$^{-1}$ per spectral element at 400 s of acquisition time with cavity-enhanced frequency comb spectroscopy, the highest demonstrated for a comb-based technique. The system comprises a frequency comb locked to a high-finesse cavity and a fast-scanning Fourier transform spectrometer with an ultra-low-noise autobalancing detector. Spectra with a signal-to-noise ratio above 1000 and a resolution of 380 MHz are acquired within a few seconds. The measured absorption lineshapes are in excellent agreement with theoretical predictions.

Continuous wave (cw) laser absorption spectroscopy is a well-established technique for quantitative measurements of various constituents in gas phase. The sensitivity of the technique can be improved either by employing modulation techniques, which shift the signal to audio [1, 2] or radio [3] frequencies, where the technical noise is reduced, or by implementing an external high-finesse enhancement cavity [4, 5], which increases the interaction length of the light with the sample, and thus the absorption signal. Successful combination of these two approaches has led to impressive shot-noise-limited absorption sensitivities [6]. The main constraint of cavity-enhanced cw techniques, however, is their inability to measure broadband spectra in short acquisition times. The optical frequency combs extend the benefits of cw-laser-based techniques to thousands of laser lines and remove the bandwidth limitation [7, 8]. The discrete spectrum of equidistant comb lines matches that of an external enhancement cavity, enabling efficient coupling of the comb light into the cavity [9, 10]. Several detection schemes have been proposed for cavity-enhanced direct frequency comb spectroscopy (CE-DFCS), but none was able to suppress the technical noise down to the quantum limit. In most previous realizations of the technique, the light transmitted through the cavity has been detected using a dispersive element and a multidetector array [11-13]. With this approach the laser frequency-to-amplitude (FM-to-AM) noise conversion caused by the narrow cavity modes can be efficiently reduced, by measuring either the cavity ringdown time [11] or integrated cavity output [12]. However, the sensitivity is limited by the technical noise in the detector array because of the low optical power in each spectrally resolved element. An alternative is to use Fourier transform spectrometry (FTS), either with a mechanical Michelson interferometer [14, 15] or with the dual comb approach [16]. The advantage of using FTS for CE-DFCS is that the spectral bandwidth of the detection system is limited only by the cavity dispersion, and not by the number of detection elements [17]. However, a photodetector and data acquisition board with a large dynamic range are necessary to obtain high signal-to-noise ratios [18]. Moreover, the use of FTS requires constant transmission through the cavity and thus the comb to be locked to the cavity, which introduces intensity noise due to FM-to-AM conversion. Here we overcome these difficulties and achieve for the first time a shot-noise-limited sensitivity in CE-DFCS by tightly locking a frequency comb to a high-finesse cavity and employing a fast-scanning FT interferometer with an ultra-low-noise autobalanced dual beam detection. Reaching the fundamental quantum limit makes frequency comb spectroscopy faster and more precise, which is crucial for modern applications such as medical diagnosis [12, 19], probing of ultracold molecules [20] or comb-based quantum information processing [21].

The system, schematically depicted in Fig. 1, is based on an Er:fiber femtosecond laser with a repetition rate of 250 MHz and emission spanning the wavelength range from 1510 to 1610 nm. The comb is locked by the Pound-Drever-Hall technique [22] to an external cavity with a length, $L$, of 60 cm and a finesse, $F$, of 8000 using error signals derived at two different wavelengths [23, 24]. The feedback with a bandwidth of 20 kHz is sent to the laser current and a fast PZT in the laser cavity, thus stabilizing both the carrier envelope offset frequency and the comb repetition frequency to the external cavity. The details of the laser locking to the cavity are outlined in the supplemental information. The PZT correction signal is further integrated and sent to a large range PZT on which the output mirror of the external cavity is mounted to compensate for long term drifts. As a result, two (groups of) comb lines are locked to their respective cavity modes, ensuring constant transmission over hours of operation. Despite the dispersion of the cavity mirrors, up to 50 nm of laser spectrum can be simultaneously transmitted through the cavity, depending on the choice of the locking points. The transmitted light is analyzed with a home-built fast-scanning FT interferometer [25]. The two output beams of the interferometer are incident on a low-noise autobalancing InGaAs photodetector based on a Hobbs design [26], optimized to reduce the nonlinear effects in the photodiodes and transistors caused by the pulsed nature of the frequency comb. The resulting interferogram is digitized with a 22 bit data acquisition board at 1 Msample/s rate. The interferogram is then resampled at the zero crossings of a high-pass filtered interferogram of a cw 780 nm external cavity diode laser (ECDL), whose beam is propagating parallel to the frequency comb beam in the spectrometer.

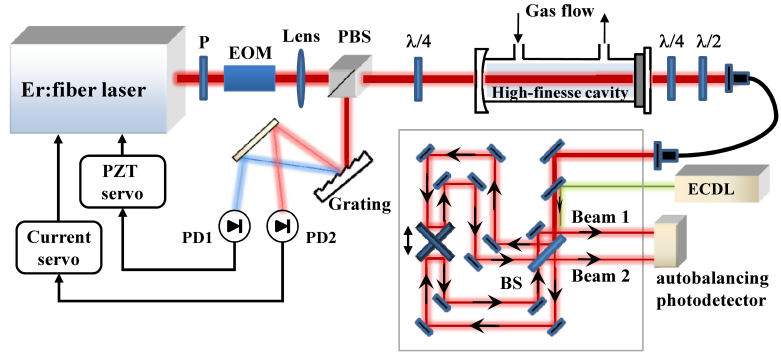

**Figure 1**. Schematic of the experimental setup. Er:fiber femtosecond laser is locked to a high-finesse optical cavity containing a gas sample. An electro-optic modulator (EOM) is used to phase modulate the comb light at 14 MHz, and the cavity reflected light is dispersed by a reflection grating and imaged on two photodetectors (PD1 and PD2) in order to create error signals at two different wavelengths. The feedback is sent to laser current controller and to a PZT inside the laser cavity. The cavity transmitted light is coupled through a polarization-maintaining fiber into a fast-scanning Fourier transform spectrometer. The two outputs of the interferometer (beams 1 and 2) are incident on two photodiodes of the autobalancing photodetector (A-PD). The beam of a cw 780 nm external cavity diode laser (ECDL), used for frequency calibration, is propagating parallel to the frequency comb beam and incident on a separate detector (not shown). P – polarizer, (P)BS – (polarizing) beam splitter cube, λ/4 - quarter wave plate, λ/2 - half wave plate.

The photocurrents from the two photodiodes in the autobalancing detector are subtracted, and the difference of the DC photocurrents is maintained at zero by a slow feedback loop. This ensures subtraction of the common mode noise [26], while the fast interferogram signal, which is out-of-phase in the two optical beams, is doubled. Figure 2(a) shows interferograms measured with a single photodetector (i.e. with beam 2 blocked, red trace) and with the autobalancing detector (blue trace); Figure 2(b) shows their frequency domain representations. Due to the averaging inside the cavity, the laser intensity noise caused by the FM-to-AM conversion is low pass filtered at frequencies above the cavity linewidth (~15 kHz, corresponding to 500 $cm^{-1}$), which is clearly visible in the spectrum displayed in red in Fig. 2(b). Proper optimization of the locking servo parameters is critical, since tighter locking results in a lower AM-to-FM conversion. The high carrier frequency of the interferogram (180 kHz, corresponding to 6500

cm$^{-1}$) places the laser spectrum in a region of lower noise, which is, however, still 2 orders of magnitude above the shot noise level. The dual beam autobalancing detection reduces the noise at this frequency by a factor of 600 and dramatically improves the signal-to-noise ratio in the spectrum, as shown in Fig. 2(c). We found that manual balanced detection (i.e. with the feedback in the autobalancing detector disabled) in the same interferometer provides at best noise reduction by a factor of 200 and is limited by the alignment of the two optical beams.

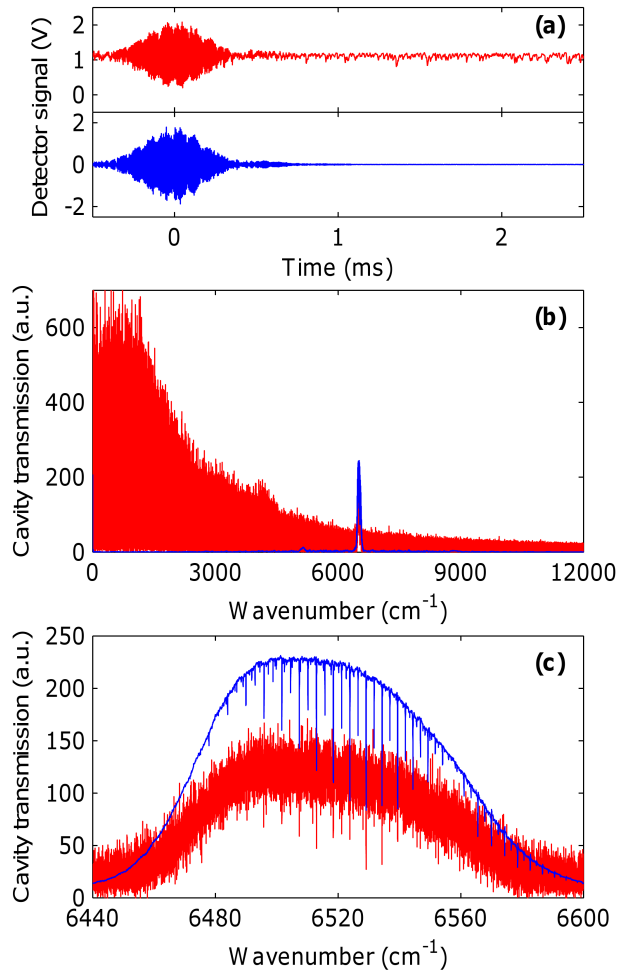

**Figure 2**. (a) The centerburst of double-sided interferograms acquired in 3 s with one photodiode (red) and with the autobalancing detector (blue) and (b) the magnitude of their Fourier transform with a resolution of 380 MHz. The cavity was filled with 5 ppm of acetylene at 150 Torr of nitrogen, the optical power detected on each photodiode of the autobalancing detector was 100 µW. The noise is reduced by a factor of 600 by the autobalancing detector at the frequency of the interferogram (180 kHz, corresponding to 6520 cm$^{-1}$), as is demonstrated in (c), where a zoom of (b) is shown.

We evaluated the sensitivity of the system at two different spectral resolutions, namely 380 MHz and 2.3 GHz. For these resolutions the acquisition time, $T$, of one interferogram is 3 s and 0.5 s, respectively (and twice as much for a normalized spectrum, i.e. the ratio of two consecutive background spectra measured with cavity filled with nitrogen); and the number of resolved spectral elements, $M$, is 10000 and 1500, respectively, in a spectrum spanning 30 nm centered around 1530 nm. The minimum detectable absorption in a single element of the spectrum, $\alpha_{min}$, was calculated as the ratio of the standard deviation of the noise in the center part of a normalized spectrum and the effective interaction length inside the cavity, given by $2FL/\pi$. Figure 3(a) shows the power dependence of the measured single element sensitivity at the two resolutions, together with the calculated shot-noise-limited sensitivity (solid lines, see supplemental information), which was confirmed by a measurement of the shot noise level on the photodetector. At a resolution of 380 MHz (red triangular markers in Fig. 3a) the system is shot-noise-limited in the 30 to 100 µW power range. The deviation from the shot noise limit at powers below 30 µW is due to the decrease of the performance of the autobalancing detector. At powers above 100 µW the intensity noise starts to dominate over the shot noise. The best shot-noise-limited single element sensitivity obtained for a 6 s acquisition time is $1.4 \times 10^{-9}$ cm$^{-1}$ at 100 µW, corresponding to $3.4 \times 10^{-11}$ cm$^{-1}$ Hz$^{-1/2}$ per spectral element (calculated as $\alpha_{min}(T/M)^{1/2}$ [25]). At a resolution of 2.3 GHz [blue square markers in Fig. 3(a)] the measured absorption noise was slightly above that given by the shot noise limit, due to the drift of the residual etalon fringes (see supplemental information). Nevertheless, a single element sensitivity of $7.6 \times 10^{-10}$ cm$^{-1}$ at 1 s was obtained, which translates to $2 \times 10^{-11}$ cm$^{-1}$ per spectral element at 1 s. The sensitivity of the spectrometer can be improved by optimizing the performance of the autobalancing detector at higher optical powers or by implementing a high bandwidth intensity servo in cavity transmission, as well as by eliminating the residual fringes.

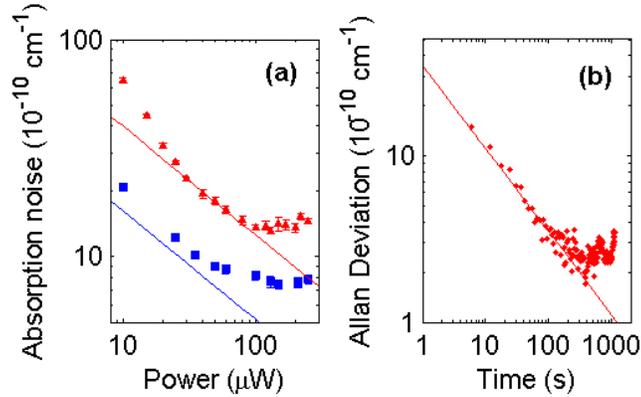

**Figure 3**. (a) Single element sensitivity in a normalized spectrum as a function of optical power incident on one photodetector at a resolution of 380 MHz (red triangular markers) and 2.3 GHz (blue square markers). Each data point is an average of 10 and 30 measurements, respectively, and the error bars are one standard error. The red and blue lines represent the corresponding shot noise limit, which is $6^{1/2}$ times lower at a resolution of 2.3 GHz because of the 6 times lower number of resolved spectral elements. (b) The Allan Deviation of a single element on the baseline between acetylene absorption lines in a spectrum measured at a resolution of 380 MHz and 100 µW of power, such as shown in Fig. 4(a). The linear fit has a $\tau^{-1/2}$ dependence characteristic for white noise up to 400 s.

To assess the long term stability of the system we measured spectra of 2 ppm of acetylene in 150 Torr of nitrogen with a resolution of 380 MHz and 100 µW of power [such as shown in Fig. 4(a)] at equal time intervals for 1 hour, and normalized them to a background spectrum recorded at the beginning of the measurement series. The Allan Deviation of the noise on a single element of the baseline between acetylene lines in the normalized spectrum is shown in Fig. 3(b). The value of the first point of the Allan Deviation (at 6 s) is the same as the shot-noise-limited sensitivity at 100 µW of power shown in Fig. 3(a). A linear fit with the slope of $3.5 \times 10^{-9}$ cm$^{-1}$ $\tau^{-1/2}$ [red line in Fig. 3(b)] confirms white noise behavior for averaging times, $\tau$, up to 7 minutes, which means that even in the presence of an absorber the system is shot-noise-limited for long averaging times. The flicker noise floor visible at longer averaging times is due to residual etalon fringes (see supplemental material). The single element sensitivity at 400 s of averaging time is $1.7 \times 10^{-10}$ cm$^{-1}$, corresponding to $1.7 \times 10^{-12}$ cm$^{-1}$ per spectral element, the lowest ever reported for direct frequency comb spectroscopy [27].

Figure 4 shows experimental spectra (in blue) of the $\nu_1 + \nu_3$ band of acetylene measured with 2 ppm of C$_2$H$_2$ in 150 Torr (panels a-c) and 630 Torr (panel d) of N$_2$. The spectrum at the lower pressure was measured in 6 s with a resolution of 380 MHz, and the spectrum at atmospheric pressure was measured in only 1 s with 2.3 GHz resolution, sufficient to resolve the pressure-broadened acetylene lines. The laser locking points were chosen at 1528 nm for the current lock and 1535 nm for the PZT lock. The spectra shown in red, inverted for clarity, are calculated using spectral line data from the HITRAN database [28] and a lineshape model presented in the supplemental material, assuming that all comb lines are on resonance with their respective cavity modes. A comparison of the theoretical model to the experimental data shows that the lineshapes of the acetylene lines are reproduced well in the parts of the spectrum close to the locking points [see the enlarged part of the spectrum in Fig. 4(c)]. At the low wavelength edge of the spectrum (away from the locking points) dispersion of the cavity mirrors introduces a mismatch between the repetition rate of the laser and the cavity free spectral range. As a result, the comb lines are located at the slope of their respective cavity modes. In the presence of acetylene, molecular dispersion shifts the cavity modes further, causing the characteristic overshoots in the absorption lineshapes, visible in the part of the spectrum enlarged in Fig. 4(b). The theoretical spectrum, shown in green (inverted), is calculated assuming a 19 kHz offset of the comb modes from the centers of their relative cavity modes and reproduces the observed lineshapes with high accuracy. The ability to correctly model molecular lineshapes is crucial for ultra-sensitive trace gas detection, since neglecting the effects of intra-cavity dispersion would introduce systematic errors in concentration determination.

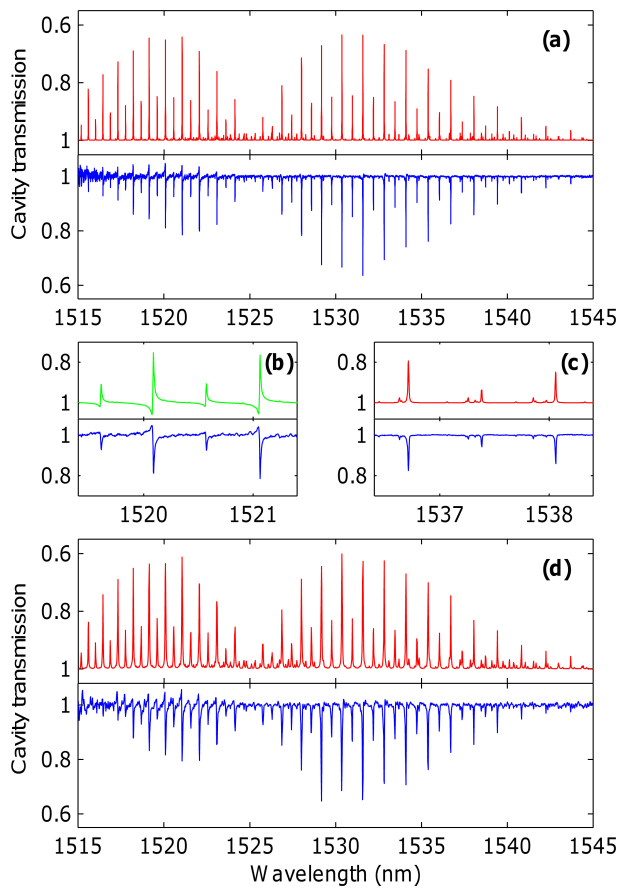

**Figure 4**. Normalized experimental (blue) and theoretical (red and green, inverted for clarity) spectra of 2 ppm of acetylene at (a)-(c) 150 Torr of $N_2$ measured in 6 s with 380 MHz resolution and at (d) 630 Torr of $N_2$ measured in 1 s with 2.3 GHz resolution. (b) and (c) show zooms of two parts of the spectrum shown in (a). (b) The slightly dispersive shape of the experimental absorption lines (blue), reproduced well in the model spectrum (green, inverted), is caused by the fact that the comb lines are offset from their respective cavity modes at wavelengths away from the laser locking points (chosen at 1528 and 1535 nm in this case). (c) At the locking points and in their vicinity, the comb lines are on resonance with the cavity modes and the absorption lines lack the dispersive part. The signal-to-noise ratio is lower in (b) than in (c) because of the lower comb power at these wavelengths.

We have demonstrated for the first time cavity-enhanced direct frequency comb spectroscopy with shot-noise-limited absorption sensitivity, the highest ever achieved with a comb-based technique. High resolution broadband spectra with a signal-to-noise ratio above 1000 are acquired within seconds, and the system provides stable operation over hours. The measured absorption lineshapes are in excellent agreement with theoretical model, which enables multiline fitting over the whole available spectral bandwidth for improved sensitivity and precision of trace species concentration determination [15]. The technique is thus ready for applications that demand ultrasensitive multispecies detection, such as breath analysis or detection of pollutants and hazardous gases. Implementing a mid-infrared frequency comb will grant access to the strong fundamental transitions of many molecules and yield extremely low concentration detection limits. Finally, quantum-noise-limited frequency comb spectroscopy can trigger new applications of the frequency comb such as quantum information processing.

**Acknowledgments**

The authors thank Terry Brown for developing the autobalancing detector and Long-Sheng Ma for useful comments on the manuscript. A. F. acknowledges a Swedish Research Council fellowship; T. B. is a Fulbright Fellow; and P. M. holds a fellowship from the Polish Ministry of Science and Higher Education. This project is supported by AFOSR, DTRA, NIST, NSF, and Agilent. The permanent address for T. B. is Institute of Physics, Bijenicka Cesta 46, Zagreb, Croatia. The permanent address for P. M. is Instytut Fizyki, Uniwersytet Mikołaja Kopernika, Torun, Poland. Current address for F. A. is NIST, 325 Broadway, Boulder, CO 80305.

**References**

[1] H. Wahlquist, J. Chem. Phys. **35**, 1708-1710 (1961).

[2] P. Kluczynski, J. Gustafsson, A. M. Lindberg, and O. Axner, Spectroc. Acta B **56**, 1277-1354 (2001).

[3] G. C. Bjorklund, Opt. Lett. **5**, 15 (1980).

[4] A. O'Keefe, and D. A. G. Deacon, Rev. Sci. Instrum. **59**, 2544-2551 (1988).

[5] B. A. Paldus, and A. A. Kachanov, Can. J. Phys. **83**, 975-999 (2005).

[6] J. Ye, L. S. Ma, and J. L. Hall, J. Opt. Soc. Am. B **15**, 6-15 (1998).

[7] T. Udem, R. Holzwarth, and T. W. Hansch, Nature **416**, 233-237 (2002).

[8] S. T. Cundiff, and J. Ye, Rev. Mod. Phys. **75**, 325-342 (2003).

[9] M. J. Thorpe, and J. Ye, Appl. Phys. B **91**, 397-414 (2008).

[10] F. Adler, M. J. Thorpe, K. C. Cossel, and J. Ye, Annu. Rev. Anal. Chem. **3**, 175-205 (2010).

[11] M. J. Thorpe, K. D. Moll, R. J. Jones, B. Safdi, and J. Ye, Science **311**, 1595-1599 (2006).

[12] M. J. Thorpe, D. Balslev-Clausen, M. S. Kirchner, and J. Ye, Opt. Express **16**, 2387-2397 (2008).

[13] C. Gohle, B. Stein, A. Schliesser, T. Udem, and T. W. Hansch, Phys. Rev. Lett. **99**, 263902 (2007).

[14] J. Mandon, G. Guelachvili, and N. Picque, Nat. Photonics **3**, 99-102 (2009).

[15] F. Adler *et al.*, Opt. Express **18**, 21861-21872 (2010).

[16] I. Coddington, W. C. Swann, and N. R. Newbury, Phys. Rev. Lett. **100**, 049901 (2008).

[17] B. Bernhardt *et al.*, Nat. Photonics **4**, 55-57 (2010).

[18] N. R. Newbury, I. Coddington, and W. Swann, Opt. Express **18**, 7929-7945 (2010).

[19] T. H. Risby, and S. F. Solga, Appl. Phys. B **85**, 421-426 (2006).

[20] D. Wang *et al.*, Phys. Rev. A **81**, 061404 (2010).

[21] N. C. Menicucci, S. T. Flammia, and O. Pfister, Phys. Rev. Lett. **101**, 130501 (2008).

[22] R. W. P. Drever *et al.*, Appl. Phys. B **31**, 97-105 (1983).

[23] R. J. Jones, and J. C. Diels, Phys. Rev. Lett. **86**, 3288-3291 (2001).

[24] R. J. Jones, I. Thomann, and J. Ye, Phys. Rev. A **69**, 051803 (2004).

[25] A. Foltynowicz *et al.*, Faraday Disc. **150**, 23-31 (2010).

[26] P. C. D. Hobbs, Appl. Opt. **36**, 903-920 (1997).

[27] The sensitivity quoted in Ref. 17, which appears to be lower than that demonstrated here, is incorrectly normalized to Hz$^{-1/2}$, as is explained in Ref. 18.

[28] L. S. Rothman *et al.*, J. Quant. Spectrosc. Radiat. Transf. **110**, 533-572 (2009).

**Supplemental information**

**Shot-noise-limited sensitivity of cavity-enhanced spectroscopy with continuous wave lasers and optical frequency combs.** The minimum detectable absorption coefficient, $\alpha_{\min}$, in shot-noise-limited cavity-enhanced absorption spectroscopy with cw lasers is given by [1]

$$\alpha_{\min}^{cw} = \sqrt{\frac{2eB}{\eta P}} \frac{\pi}{kFL}, \qquad (1)$$

where $e$ is the electronic charge, $B$ is the electronic bandwidth, $\eta$ the detector responsivity, $P$ the power incident on the detector, $F$ the cavity finesse, $L$ the cavity length and $k$ takes a value between 1 and 2, depending on the method of coupling of the laser to the cavity [2]. In the case of a cw laser which is scanned over $M$ spectral elements in time $T$ the effective bandwidth is given by $B = M/T$ and equation (1) becomes

$$\alpha_{\min}^{cw} = \sqrt{\frac{2eM}{\eta PT}} \frac{\pi}{kFL}. \qquad (2)$$

The same equation is valid for cavity-enhanced frequency comb spectroscopy employing a dispersive element and a detector array, where $M$ spectral elements are acquired simultaneously in time $T$, but the power in each spectral element is only $P/M$.

The minimum detectable absorption coefficient, $\alpha_{\min}^{FC-FTS}$, in a single element of a normalized spectrum for shot-noise-limited cavity-enhanced frequency comb FTS (assuming 100% modulation depth on the interferogram) is given by [3, 4]

$$\alpha_{\min}^{FC-FTS} = \sqrt{\frac{2eB}{\eta P}} \frac{M}{\sqrt{N}} \frac{\pi}{2FL}. \qquad (3)$$

Here $B$ is the effective electronic bandwidth in the time domain given by $N/T$, where $N$ is the number of points in the interferogram acquired in time $T$. The ratio $\sqrt{N}/M$ relates the signal-to-noise ratio in the time and frequency domains. Thus, assuming the same total optical power, acquisition time and number of spectral elements, the sensitivity of a broadband measurement with a detector array and a cw measurement is the same and better by $\sqrt{M}$ than that of frequency comb based FTS in the shot-noise-limited case. However, *i*) the shot noise limit is currently impossible to obtain with a detector array because of the low power in each spectral component; *ii*) acquisition of thousands of spectral elements with a cw laser and a cavity over a spectral range comparable to that accessible with a frequency comb requires in practice much longer measurement times.

**Two point locking of an optical frequency comb to a high finesse cavity.** In order to efficiently couple frequency comb light into a high finesse cavity the repetition rate ($f_r$) of the comb has to be equal to the cavity free spectral range (FSR) and the carrier envelope offset frequency ($f_0$) has to be adjusted properly to ensure that comb modes line up with that of an external cavity [2, 5]. Achieving a good match of the comb and cavity spectra is however usually possible only over a limited wavelength range, determined by the dispersion of the cavity mirrors, which causes the cavity FSR to vary with optical frequency. Moreover, the linewidth of frequency comb lines should be narrower than that of the cavity in order to minimize the amplitude to frequency noise conversion. The $f_r$ and $f_0$ of the particular Er:fiber laser used in this experiment could be controlled by changing the oscillator pump diode current or tuning the PZT in the oscillator cavity. The control of the two frequencies was not orthogonal [6]; the $f_0$ was changing by -1 MHz per mA of pump current and by -0.4 MHz per V on the intracavity PZT; the $f_r$ was changing by 13 Hz/mA and 4 Hz/V, respectively. We implemented a two point locking scheme based on that described in Refs [7, 8], in which error signals derived at different spectral regions provide information about the average frequency of the comb and the repetition frequency. In order to ensure a wide transmitted bandwidth and narrow linewidth of each comb line a modified locking scheme was used here. The light reflected from the cavity was dispersed with a reflection grating and incident on two photodiodes, each detecting approximately 1 nm of bandwidth of light resonant with the cavity around two locking points separated by a few nm. Two error signals were obtained using the standard Pound-Drever-Hall technique [9]. Each error signal was used to lock a group of comb lines to their respective cavity modes by controlling both $f_r$ or $f_0$ via feedback to either pump current or intracavity PZT. The wavelengths of the locking points were chosen so as to maximize the transmission through the cavity at the wavelength range of interest (in this experiment mostly 1515-1545 nm, where strongest acetylene absorption occurs). In general the locking points should be as far apart as possible within the emission spectrum of the laser and in the range where the cavity FSR is relatively constant (so that locking to one of the points does not introduce a significant offset in the error signal at the other locking point). The bandwidth of both feedback loops was 20 kHz, which resulted in the residual frequency jitter of the comb lines with respect to the cavity modes of less than 50% of the cavity mode linewidth (half width at half maximum, equal to 15 kHz). Away from the locking points the comb modes walk off the centers of their respective cavity modes because of the change of the cavity FSR, until the relative shift becomes significantly larger than the cavity mode linewidth and the comb modes are not transmitted through the cavity at all.

**Theoretical model of molecular spectrum in cavity transmission.** The optical frequency ($\nu$) dependent intensity of light transmitted through the cavity in the presence of absorbing gas, $I_t(\nu)$, normalized to the intensity in the absence of the analyte, $I_0(\nu)$, was calculated using

$$\frac{I_t(\nu)}{I_0(\nu)} = \frac{t^2(\nu)e^{-2\delta(\nu)L}}{1-r^2(\nu)e^{-4\delta(\nu)L}-2r(\nu)e^{-2\delta(\nu)L}\cos\left[2\phi(\nu)L+\varphi(\nu)\right]}, \quad (4)$$

where $t(\nu)$ and $r(\nu)$ are the frequency-dependent transmission and reflection coefficients of the cavity mirrors, respectively, $L$ is the cavity length [cm], $\varphi(\nu)$ is the round trip phase shift in the cavity, given by $4\pi\nu n L/c$, where $n$ is the refractive index and $c$ the speed of light; and where $\delta(\nu)$ and $\phi(\nu)$ are the single pass attenuation and phase shift of the electric field inside the cavity due to the analyte, given by

$$\delta(\nu) = \frac{Sn_A}{2}\mathrm{Re}\,\chi(\nu), \quad (5)$$

$$\phi(\nu) = \frac{Sn_A}{2}\mathrm{Im}\,\chi(\nu). \quad (6)$$

Here $S$ is the molecular linestrength in [cm$^{-1}$/molecule/cm$^{-2}$], $n_A$ the density of absorbers [molecules/cm$^3$], and $\chi(\nu)$ is the complex lineshape function [cm], in the pressure broadened regime given by the Voigt profile (i.e. error function of a complex argument). Note that absorption is equal to twice the attenuation of the electric field, i.e. $\alpha(\nu)=2\delta(\nu)$.

For highly reflective mirrors the reflectivity is given by $r(\nu)=1-\pi/F(\nu)$, where $F(\nu)$ is the cavity finesse, and it is assumed that $r(\nu)+t(\nu)=1$. The frequency-dependent cavity finesse was measured using cavity ringdown with the Er:fiber femtosecond laser and a monochromator behind the cavity.

When a comb line is on resonance with its respective cavity mode, the round trip intracavity phase shift is equal to $2q\pi$, where $q$ is an integer mode number. When a comb line is locked to the slope of the cavity mode instead, the intracavity phase shift differs from a multiple of $2\pi$. In the particular spectrum shown in Fig. 4(b) in the main text, the intracavity phase shift relative to the resonance condition is equal to $-5\times10^{-4}$ rad. This corresponds to a 19 kHz offset between the position of the comb modes and their respective cavity modes, which is given by $[\varphi(\nu)-2q\pi]/(2\pi FSR)$, where $FSR = c/(2L)$ is the free spectral range of the cavity, equal to 250 MHz. The cavity finesse at 1520 nm is 6600, implying that the 19 kHz detuning from the center of the cavity mode is equal to roughly one half width of the cavity mode. The agreement of this theoretical model with the observed lineshapes is excellent, as is demonstrated in Fig. 4 in the main text.

**Etalon fringes in the cavity transmitted spectrum**. The sensitivity of the spectrometer at times longer than 7 minutes was limited by flicker noise originating from residual etalon fringes, as is stated in the vicinity of Fig. 2(c) in the main text. The optical elements were tilted, whenever possible, in order to avoid multiple reflections (etalon effects) within and between these components. Polarization of the light entering the EOM and the polarization-maintaining fiber after the cavity was carefully aligned to be linear along the e-axis of the EOM and the fiber connector key, respectively, to avoid etalon effects. We were able to identify and then eliminate all but two relatively large etalon fringes in the spectrum, one from reflection within the cavity mirrors, the other from within a variable neutral density filter, which was used to obtain an optimum power balance on the autobalancing detector. Due to a slight drift of the cavity transmitted spectrum, these etalons caused small ripples on the normalized spectra. In order to remove them from acetylene spectra, a sum of sinusoidal fringes was fitted to the points between acetylene absorption lines, interpolated at the points where absorption occurs and subtracted from the final spectrum. Note that in order to minimize the influence of reflections between optical components (mainly the cavity mirrors and fiber collimator etc.), these components can be placed at distances such that the free spectral range of the etalon is below the resolution of the spectrometer. In other words, if the length of the etalon is longer than the maximum optical path difference in the FTS, the reflected pulse will not be seen. Care should be taken, though, to avoid folding of reflections from optical components placed at distances longer than the cavity length.

**References**

[1] J. Ye, L. S. Ma, and J. L. Hall, J. Opt. Soc. Am. B **15**, 6-15 (1998).

[2] M. J. Thorpe, and J. Ye, Appl. Phys. B **91**, 397-414 (2008).

[3] N. R. Newbury, I. Coddington, and W. Swann, Opt. Express **18**, 7929-7945 (2010).

[4] A. Foltynowicz *et al.*, Faraday Disc. **150**, 23-31 (2010).

[5] F. Adler, M. J. Thorpe, K. C. Cossel, and J. Ye, Annu. Rev. Anal. Chem. **3**, 175-205 (2010).

[6] N. R. Newbury, and W. C. Swann, J. Opt. Soc. Am. B **24**, 1756-1770 (2007).

[7] R. J. Jones, and J. C. Diels, Phys. Rev. Lett. **86**, 3288-3291 (2001).

[8] R. J. Jones, I. Thomann, and J. Ye, Phys. Rev. A **69**, 051803 (2004).

[9] R. W. P. Drever *et al.*, Appl. Phys. B **31**, 97-105 (1983).